\documentclass[preprint,12pt]{elsarticle}

\usepackage{amssymb}
\usepackage{epsfig}
\usepackage{epsf}
\usepackage{booktabs}
\usepackage{amssymb}
\usepackage{multirow}
\usepackage{mathrsfs}
\usepackage{longtable}
\usepackage{lscape}
\usepackage{tabularx}
\usepackage{subfigure}
\usepackage{amsmath}
\usepackage[mathlines]{lineno}

\usepackage{fancyhdr}

\newdefinition{definition}{Definition}
\newdefinition{example}{Example}
\newdefinition{remark}{Remark}
\newdefinition{theorem}{Theorem}

\journal{Journal of Theoretical Biology}

\begin{document}

\begin{frontmatter}

\title{A belief-based evolutionarily stable strategy}

\author[SWU]{Xinyang Deng\fnref{eqw}}
\author[HKBU,HKB]{Zhen Wang\fnref{eqw}}
\author[CQS,DBI]{Qi Liu}
\author[SWU,NPU,SOE]{Yong Deng\corref{COR}}
\ead{prof.deng@hotmail.com; yongdeng@nwpu.edu.cn}
\author[SOE]{Sankaran Mahadevan}

\cortext[COR]{Corresponding author: Yong Deng, School of Computer and Information Science, Southwest University, Chongqing 400715, China.}
\fntext[eqw]{These authors contributed equally to this work.}
\address[SWU]{School of Computer and Information Science, Southwest University, Chongqing, 400715, China}
\address[HKBU]{Department of Physics, Hong Kong Baptist University, Kowloon Tong, Hong Kong, China}
\address[HKB]{Center for Nonlinear Studies, Beijing-Hong Kong-Singapore Joint Center for Nonlinear and Complex systems (Hong Kong), and Institute of Computational and Theoretical Studies, Hong Kong Baptist University, Kowloon Tong, Hong Kong, China}
\address[CQS]{Center for Quantitative Sciences, Vanderbilt University School of Medicine, Nashville, TN, 37232, USA}
\address[DBI]{Department of Biomedical Informatics, Vanderbilt University School of Medicine, Nashville, TN, 37232, USA}
\address[NPU]{School of Automation, Northwestern Polytechnical University, Xi'an, Shaanxi, 710072, China}
\address[SOE]{School of Engineering, Vanderbilt University, Nashville, TN, 37235, USA}

\begin{abstract}
As an equilibrium refinement of the Nash equilibrium, evolutionarily stable strategy (ESS) is a key concept in evolutionary game theory and has attracted growing interest. An ESS can be either a pure strategy or a mixed strategy. Even though the randomness is allowed in mixed strategy, the selection probability of pure strategy in a mixed strategy may fluctuate due to the impact of many factors. The fluctuation can lead to more uncertainty. In this paper, such uncertainty involved in mixed strategy has been further taken into consideration: a belief strategy is proposed in terms of Dempster-Shafer evidence theory. Furthermore, based on the proposed belief strategy, a belief-based ESS has been developed. The belief strategy and belief-based ESS can reduce to the mixed strategy and mixed ESS, which provide more realistic and powerful tools to describe interactions among agents.
\end{abstract}

\end{frontmatter}

%
%

\section{Introduction}
Game theory \cite{von2007theory,myerson2013game} provides an effective mathematical framework to explain and study the interactions among individuals. In many situations, the preferences, aims, and goals of participating individuals are potentially in conflict \cite{myerson1991game}. A canonical example is the prisoner's dilemma game \cite{poundstone2011prisoner}, which exhibits an apparent social dilemma that human cooperation disappears when there exists a conflict between individual and collective rationality. Due to its significant advantages of depicting the essence underlying many phenomena in nature and society, game theory has been widely used in scientific disciplines from economics, psychology to biology, as well as operational research and political science.

Recently, with the ample introduction of temporal dynamics and spatial topology, traditional theory has been elevated to a new flat: evolutionary game theory \cite{lewontin1961evolution,hamilton1967extraordinary,smith1973the,smith1982evolution,hammerstein1994game,dugatkin1998game,masuda_pla03,huttegger2013methodology,nowak2004evolutionary}, which provides a paradigmatic framework to study the evolution of cooperation within population dynamics \cite{hofbauer1998evolutionary,santos2006evolutionary,axelrod2006evolution,szabo2007evolutionary,nowak2006evolutionary,santos2012dynamics}. Along this research line, the mechanisms of promoting the emergence of cooperative behaviors have been greatly proposed \cite{nowak2006five,zimmermann2005cooperation,santos2005scale,
gomez2007dynamical,hauert_jtb06b,wang2013insight,ohtsuki2006simple,QingJinSR20144095,
sinatra2009ultimatum,genki_sr,wang2013impact,szolnoki2012wisdom,
li2014comprehensive,wang2014rewarding}. Typical examples include the the mobility of players \cite{vainstein_jtb07,meloni_pre09,xia12,jiang_ll_pre10,helbing_pnas09}, heterogeneous activity \cite{szolnoki_epl07,zhen_plos10}, spatial structured population \cite{roca_prl06,LazaroPNAS12,traulsen_pre04,cao_xb_pa10,wu_zx_pre06}, and coevolutionary selection of dynamical rules \cite{jun_pre12,perc_bs10}, to name but a few. In spite of plentiful achievements, a basic conception, evolutionarily stable strategy (ESS), which was first proposed by Smith and Price \cite{smith1973the} and further explained in \cite{smith1982evolution,maynard1974theory}, always attracts the firm attention from theoretical and experimental viewpoints \cite{taylor1978evolutionary,taylor2004evolutionary,nowak2004emergence,nowak2006evolutionary,tarnita2009mutation,shakarian2012review}. An ESS, an equilibrium refinement of the Nash equilibrium \cite{nash1950equilibrium}, can be regarded as a solution of one specific game, which is self enforcing and where no player can gain benefit by unilaterally deviating from it. At variance with Nash equilibrium, ESS is evolutionarily stable, and it can be either pure or mixed. Previous study \cite{smith1982evolution} demonstrated that a game with two pure strategies always has an ESS, despite it is either a pure ESS or a mixed ESS in a infinite population.

The mixed strategy usually reflects the randomness of strategies. For example, in a given game with two pure strategies, these two pure strategies can be represented by $s_1$, $s_2$. In this line, the mixed strategy can be expressed as $I = Ps_1 + (1-P)s_2$, where $P$ determines the probability of strategy $s_1$ to be selected and takes a value within the interval $[0, 1]$. However, due to the impact of many factors, the selection probability of pure strategy keeps fluctuating in a range $[P^-, P^+]$ rather than being constant. While this disturbance can be caused by environmental noise, agent's rationality degree and other factors, which  gives rise to more uncertainty than that of mixed strategy.

Aiming to represent the uncertainty, one novel strategy, belief strategy, is proposed in this paper. Interestingly, the belief strategy is based upon the Dempster-Shafer evidence theory \cite{Dempster1967,Shafer1976}, which is a tool of expressing uncertainty and exploring questions under the uncertain environment \cite{denoeux2004evclus,denoeux2013maximum,XinyangDAHP,Kang2012,Deng2013Environment,yang2013evidential,DengTOPPER2013,yager2013decision}. Then, the proposed belief strategy is used to extend the ESS, namely, the belief strategy is a generalization of mixed strategy and the belief-based ESS is a generalization of mixed ESS. This setup provides more realistic and powerful frameworks to describe interactions among agents. The remainder of this paper is organized as follows.  We will first describe the Hawk-Dove game, evolutionarily stable strategy, and Dempster-Shafer evidence theory; subsequently, we will present the proposed belief strategy and belief-based ESS; and finally we will summarize our conclusions.

\section{Preliminaries}\label{Sect2}
\subsection{Hawk-Dove game and evolutionarily stable strategy (ESS)}
\begin{figure}[htbp]
\begin{center}
\psfig{file=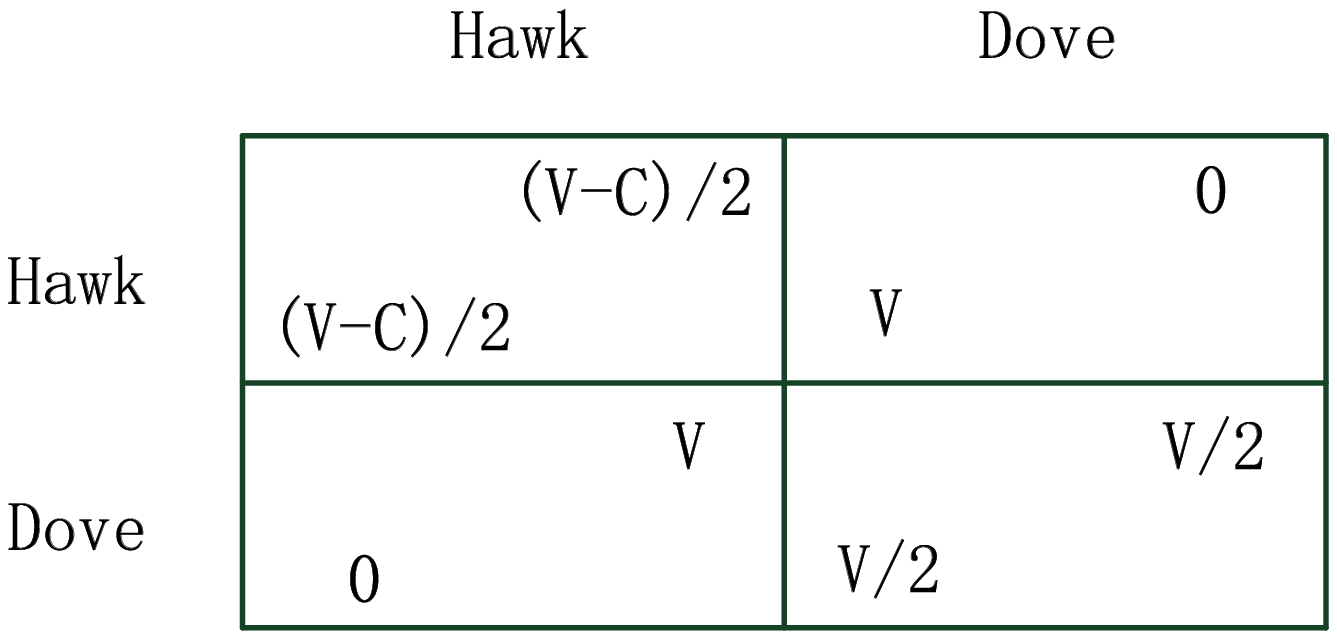,scale=0.65,angle=-90}
\caption{The payoff matrix of Hawk-Dove game}\label{HDGame}
\end{center}
\end{figure}

Hawk-Dove game \cite{smith1973the,smith1982evolution,szabo2007evolutionary} is a simple, paradigmatic model to simulate the competition between animals. Assume there is a population of animals, in which each individual aggressiveness is different during the interaction with others. Accordingly, their behaviors can be divided into two types: the aggressive type and the cooperative type. The aggressive type corresponds to strategy ``Hawk" (H), the cooperative type is associated with strategy ``Dove" (D). Within each interaction, two animals meet and compete for a resource $V$ ($V>0$). When two Hawks meet, they will fight so that both of them have the opportunity to get $(V-C)/2$, where $C$ is the cost of injury in the fight. When two Doves meet, they will share the resource, which means each individual obtains $V/2$. If, however, a Hawk meets a Dove, the former will fight and the latter can only escape. As a result, the Hawk obtains the entire resource without any cost of injury, the Dove is left with nothing. In this sense, The payoff matrix of Hawk-Dove game is shown in Figure \ref{HDGame}.

Evolutionarily stable strategy (ESS) is a key concept in evolutionary game theory. According to its definition \cite{smith1973the}, in a given environment an ESS is such a strategy that can not be invaded by any other alternative strategy which is initially rare. The condition required by an ESS can be formulated as \cite{smith1982evolution,maynard1974theory}:
\begin{equation}\label{ESScondition1}
E(S,S) > E(T,S),
\end{equation} or
\begin{equation}\label{ESScondition2}
E(S,S) = E(T,S), \qquad and \qquad E(S,T) > E(T,T),
\end{equation}
for all $T \ne S$, where strategy $S$ is an ESS, $T$ is an alternative strategy, and $E(T,S)$ is the payoff of strategy $T$ playing against strategy $S$.

The conditions given in above equations (namely, Eqs.(\ref{ESScondition1}) and (\ref{ESScondition2})) are on the basis of these assumptions including infinite population, asexual inheritance, complete mixing, pairwise and symmetric contests. If the evolutionarily stable strategy (ESS) $S$ is a pure strategy, $S$ is called a pure ESS. On the contrary, once $S$ is a mixed strategy, $S$ becomes the so-called  mixed ESS. In \cite{smith1982evolution}, it has been proven that a game with two pure strategies always has an ESS (pure ESS or mixed ESS). Take the Hawk-Dove game as an example. In that game, pure strategy $H$ is an ESS if $V > C$ because $E(H,H) > E(D,H)$. Conversely, if $V < C$, the ESS of Hawk-Dove game is a mixed strategy.

While among the mixed ESS, the Bishop-Canning theorem \cite{bishop1978generalized} can provide great help. Herein, a statement given by Smith \cite{smith1982evolution} is directly adopted to display the Bishop-Canning theorem.

\textbf{Bishop-Canning theorem:} If $I$ is a mixed ESS with support $a$, $b$, $c$, $\cdots$, then
\begin{equation}
E(a,I) = E(b,I) = \cdots = E(I,I),
\end{equation}
where $a$, $b$, $c$ $\cdots$ are said to be the ``support" of $I$ if these pure strategies are played with non-zero probability in the mixed strategy.

Based on this theorem, the mixed ESS of Hawk-Dove game, denoted by $I = PH + (1 - P)D$, where $P$ is the probability choosing strategy $H$, can be expressed as
\begin{equation}\label{EHIEDI}
E(H,I) = E(D,I).
\end{equation}
Extending Eq.(\ref{EHIEDI}), we get
\begin{equation}
PE(H,H) + (1 - P)E(H,D) = PE(D,H) + (1 - P)E(D,D),
\end{equation}
\begin{equation}
P \times \frac{{V - C}}{2} + (1 - P) \times V = P \times 0 + (1 - P) \times \frac{V}{2},
\end{equation}
namely,
\begin{equation}
P = \frac{V}{C}.
\end{equation}
Hence, the mixed ESS is $I = \frac{V}{C}H + (1 - \frac{V}{C})D$. It is easy to verify that the condition displayed in Eq.(\ref{ESScondition2}) has been meet in $I$. The mixed strategy $I$ is stable against invasion.

\subsection{Dempster-Shafer evidence theory}\label{SectDempster}
Dempster-Shafer evidence theory \cite{Dempster1967,Shafer1976}, also called Dempster-Shafer theory or evidence theory, has been first proposed by Dempster \cite{Dempster1967} and then developed by Shafer \cite{Shafer1976}. This theory needs weaker conditions than the Bayesian theory of probability, so it is often regarded as an extension of the Bayesian theory. As a theory of reasoning under the uncertain environment, Dempster-Shafer theory has an advantage of directly expressing the ``uncertainty" by assigning the probability to the subsets of the set composed of multiple objects, rather than to each of the individual objects. The probability assigned to each subset is limited by a lower bound and an upper bound, which respectively measure the total belief and the total plausibility for the objects in the subset. For the simplicity of explanation, a few basic concepts are introduced as follows.

Let $\Omega$ be a set of mutually exclusive and collectively exhaustive events, indicated by
\begin{equation}
\Omega  = \{ \theta_1 ,\theta_2 , \cdots ,\theta_i , \cdots ,\theta_N \},
\end{equation}
where the set $\Omega$ is called a frame of discernment. The power set of $\Omega$ is indicated by $2^\Omega$, namely
\begin{equation}
2^\Omega   = \{ \emptyset ,\{ \theta_1 \} , \cdots ,\{ \theta_N \} ,\{ \theta_1
,\theta_2 \} , \cdots ,\{ \theta_1 ,\theta_2 , \cdots ,\theta_i \} , \cdots ,\Omega \}.
\end{equation}
The elements of $2^\Omega$ or subset of $\Omega$ are called propositions. For a frame of discernment $\Omega = \{ \theta_1 ,\theta_2 , \cdots, \theta_N \}$, a mass function is a mapping $m$ from  $2^\Omega$ to $[0,1]$, formally defined by:
\begin{equation}
m: \quad 2^\Omega \to [0,1],
\end{equation}
which satisfies the following condition:
\begin{eqnarray}
m(\emptyset ) = 0 \quad and \quad \sum\limits_{A \in 2^\Omega }
{m(A) = 1},
\end{eqnarray}
where a mass function is also called a belief function or a basic probability assignment (BPA). The assigned basic probability number $m(A)$ measures the belief being exactly assigned to proposition $A$ and represents how strongly the evidence supports $A$.

\begin{figure}[htbp]
\begin{center}
\psfig{file=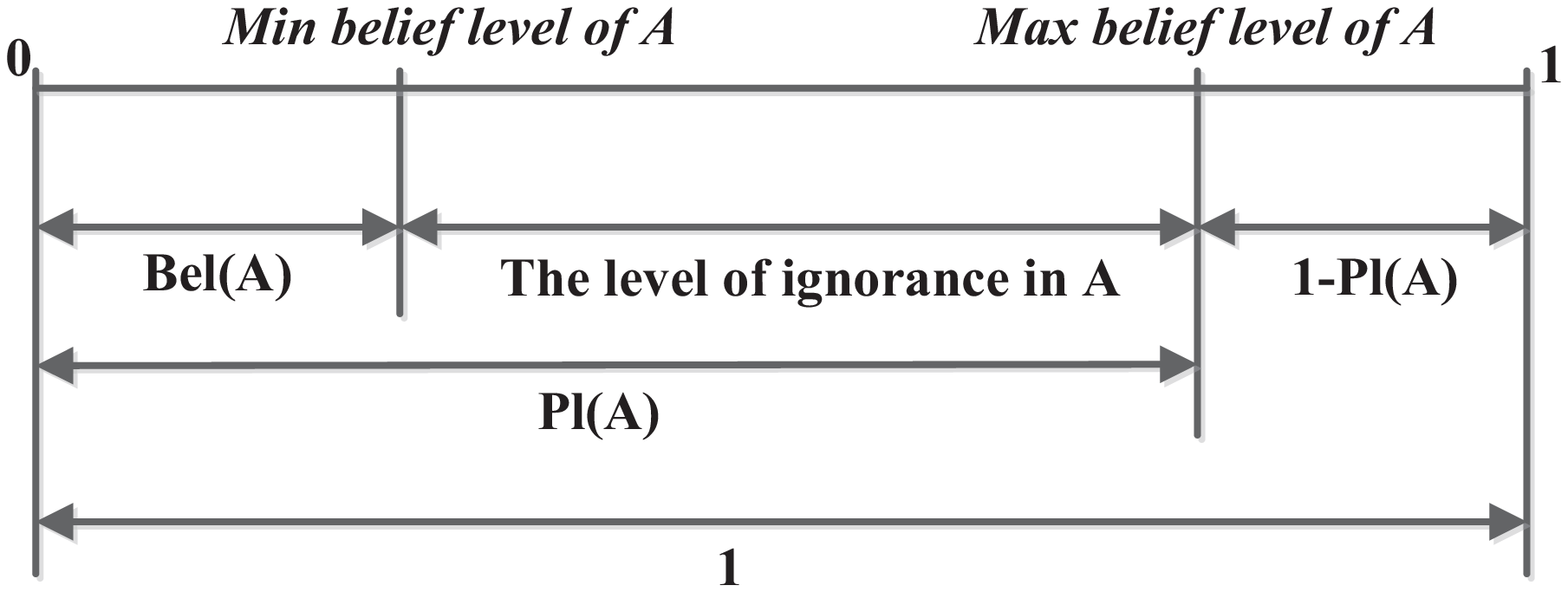,scale=0.60,angle=-90} \caption{The uncertainty expressed by belief and plausibility}\label{BeliefStrcture}
\end{center}
\end{figure}

Given a belief function $m$, we can calculate the associated belief measure and plausibility measure, indicated by $Bel$ function and $Pl$ function, respectively. For a proposition $A \subseteq \Omega$, the belief function $Bel\;:\;2^\Omega   \to [0,1]$ is defined as
\begin{equation}\label{BelFunction}
Bel(A) = \sum\limits_{B \subseteq A} {m(B)}.
\end{equation}
The plausibility function $Pl\;:\;2^\Omega   \to [0,1]$ is defined as
\begin{equation}\label{PlFunction}
Pl(A) = 1 - Bel(\bar A) = \sum\limits_{B \cap A \ne \emptyset }
{m(B)},
\end{equation}
where $\bar A = \Omega  - A$. Obviously, if $Pl(A) \ge Bel(A)$, $Bel$ and $Pl$ are the lower limit function and upper limit function of the probability to which proposition $A$ is supported, respectively. According to Shafer's explanation \cite{Shafer1976}, the difference between the belief and the plausibility of a proposition $A$ expresses the ignorance of the assessment for the proposition $A$. The uncertainty expressed by belief and plausibility is shown in Figure \ref{BeliefStrcture}.

\section{Proposed belief-based evolutionarily stable strategy}\label{Sect3}
In this section, firstly, a belief strategy is proposed based on Dempster-Shafer evidence theory, which extends the concept of mixed strategy. Secondly, in terms of the proposed belief strategy, a belief-based ESS is developed.

\subsection{Belief strategy}
As above mentioned, in game theory the strategies can be divided two types: (i) pure strategy, such as $H$ and $D$ in the Hawk-Dove game; (ii) mixed strategy, for instance $I = PH + (1 - P)D$. As for the mixed strategy, the parameter $P$ determines the selection probability of the given pure strategy. However, due to the impact of many factors, such as, environmental noise and individual rationality degree, the selection probability of pure strategy is not constant, but changes with a disturbance. Such disturbance induces the probability $P$ fluctuating in an interval $[P^-, P^+]$. In order to express such uncertainty, in this paper a new strategy type, belief strategy, is proposed based on Dempster-Shafer evidence theory. The definition of belief strategy is given as below.

\textbf{Definition of belief strategy:} Let $S = \{s_1, s_2, \cdots, s_n\}$ be the set of all pure strategies in a game, a belief strategy is a mapping $J$ from $2^S$ to $[0, 1]$, formally defined by
\begin{equation}
J: \quad 2^S \to [0,1],
\end{equation}
which satisfies
\begin{eqnarray}
J(\emptyset ) = 0 \quad and \quad \sum\limits_{A \in 2^S }{J(A) = 1}
\end{eqnarray}

If $J(A) > 0$, $A$ is called a support of belief strategy $J$. Essentially, a belief strategy can be expressed by one belief function. Take the Hawk-Dove game as an example. Assume there is an individual who adopts strategy $H$ with a probability $a$ and adopts strategy $D$ with a probability $b$, where $a,b \ge 0$ and $a + b \le 1$. The reminder $(1-a-b)$ is indistinguishable so that it is assigned to the mixture of $H$ and ${D}$, namely $\{H, D\}$. So the individual strategy is indicated by
\begin{equation}\label{BeliefStrategyJMass}
\left\{ \begin{array}{l}
 J(H) = a, \\
 J(D) = b, \\
 J(H,D) = 1 - a - b. \\
 \end{array} \right.
\end{equation}

By means of the definitions of belief function and plausibility function, shown in Eqs.(\ref{BelFunction}) and (\ref{PlFunction}), the lower limit and upper limit of each pure strategy's selection probability can be derived as:
\begin{equation}
Bel(H) = m(H) = a, \quad Pl(H) = m(H) + m(H,D) = 1 - b.
\end{equation}
\begin{equation}
Bel(D) = m(D) = b, \quad Pl(D) = m(D) + m(H,D) = 1 - a.
\end{equation}

According to the definitions of $Bel$ and $Pl$ functions, the following relation is satisfied,
\begin{equation}
\left\{ \begin{array}{l}
 Bel(D) = 1 - Pl(H) ,\\
 Pl(D) = 1 - Bel(H) .\\
\end{array} \right.
\end{equation}

Hence, in contrast with mixed strategy, the belief strategy $J$ shown in Eq.(\ref{BeliefStrategyJMass}) can also represented as
\begin{equation}
J = [Bel(H),Pl(H)] \otimes H + [1 - Pl(H),1 - Bel(H)] \otimes D,
\end{equation}
or
\begin{equation}
J = tH + (1-t)D,  \qquad t \in [Bel(H), Pl(H)].
\end{equation}

Conceptually, the mixed strategy is a generalization of pure strategy, the belief strategy is a generalization of mixed strategy. If the set of supports of belief strategy $J$ only consists of single pure strategy, $J$ is reduced to a mixed strategy.

\subsection{Belief-based ESS}
Based on the above belief strategy, the belief-based ESS can been proposed. Similar to the conditions of classically pure ESS and mixed ESS, a belief strategy $J$ can become the belief-based ESS, which is stable against the invasion of alternative strategy $T$, only if
\begin{equation}\label{BESScondition1}
\mathbb{E}\left[ {J,J} \right] > \mathbb{E}\left[ {T,J} \right],
\end{equation} or
\begin{equation}\label{BESScondition2}
\mathbb{E}\left[J,J\right] = \mathbb{E}\left[ T,J\right], \qquad and \qquad \mathbb{E}\left[ J,T\right] > \mathbb{E}\left[T,T\right],
\end{equation}
for all $T \ne J$, $\mathbb{E}\left[ J,T\right]$ is the expected payoff of strategy $J$ playing against strategy $T$. The above conditions are also just suitable for infinite population, pairwise and symmetric contests.

In order to find the belief-based ESS, it is necessary to calculate the lower limit and upper limit of selection probability of each pure strategy, whereat the Bishop-Canning theorem is also used. Take the Hawk-Dove game as an example with $V < C$. The belief strategy is $J = tH + (1-t)D$, $t \in [Bel(H), Pl(H)]$. Herein we assume that $t$ is uniformly distributed in the interval $[Bel(H), Pl(H)]$, and its probability density function is displayed as follows,
\begin{equation}
f(t) = \left\{ \begin{gathered}
  \frac{1}
{{Pl(H) - Bel(H)}},\quad \quad Bel(H) \leqslant t \leqslant Pl(H), \hfill \\
  \qquad \qquad 0,\qquad \quad \qquad \qquad otherwise. \hfill
\end{gathered}  \right.
\end{equation}

Due to $H$ and $D$ are the supports of $J$, we get,
\[
\begin{gathered}
  \mathbb{E}\left[ {H,J} \right] = \int_{ - \infty }^{ + \infty } {\left[ {tE(H,H) + (1 - t)E(H,D)} \right]f(t)dt}  \hfill \\
  \quad \quad  \;\; = \int_{Bel(H)}^{Pl(H)} {\frac{{tE(H,H) + (1 - t)E(H,D)}}
{{Pl(H) - Bel(H)}}dt}  \hfill \\
  \quad \quad  \;\; = \int_{Bel(H)}^{Pl(H)} {\frac{{t\left[ {E(H,H) - E(H,D)} \right]}}
{{Pl(H) - Bel(H)}}dt}  + \int_{Bel(H)}^{Pl(H)} {\frac{{E(H,D)}}
{{Pl(H) - Bel(H)}}dt}  \hfill \\
  \quad \quad  \;\; = \frac{{E(H,H) - E(H,D)}}
{{Pl(H) - Bel(H)}} \cdot \left. \frac{{t^2 }}
{2} \right|_{Bel(H)}^{Pl(H)}  + \frac{{E(H,D)}}
{{Pl(H) - Bel(H)}} \cdot \left. t \right|_{Bel(H)}^{Pl(H)}  \hfill \\
  \quad \quad  \;\; = \frac{{E(H,H) - E(H,D)}}
{{Pl(H) - Bel(H)}}\left( {\frac{{\left[ {Pl(H)} \right]^2 }}
{2} - \frac{{\left[ {Bel(H)} \right]^2 }}
{2}} \right) + \frac{{E(H,D)}}
{{Pl(H) - Bel(H)}}\left[ {Pl(H) - Bel(H)} \right] \hfill \\
  \quad \quad  \;\; = \left[ {E(H,H) - E(H,D)} \right]\frac{{Pl(H) + Bel(H)}}
{2} + E(H,D) \hfill \\
  \quad \quad  \;\; = \left( {\frac{{V - C}}
{2} - V} \right)\frac{{Pl(H) + Bel(H)}}
{2} + V \hfill \\
  \quad \quad  \;\; = V - \frac{{V + C}}
{2} \cdot \frac{{Pl(H) + Bel(H)}}
{2} \hfill \\.
\end{gathered}
\]
Similarly,
\[
\begin{gathered}
\mathbb{E}\left[ {D,J} \right] = \int_{ - \infty }^{ + \infty } {\left[ {tE(D,H) + (1 - t)E(D,D)} \right]f(t)dt}  \hfill \\
  \quad \quad  \;\; = \int_{Bel(H)}^{Pl(H)} {\frac{{tE(D,H) + (1 - t)E(D,D)}}
{{Pl(H) - Bel(H)}}dt}  \hfill \\
  \quad \quad  \;\; = \int_{Bel(H)}^{Pl(H)} {\frac{{t\left[ {E(D,H) - E(D,D)} \right]}}
{{Pl(H) - Bel(H)}}dt}  + \int_{Bel(H)}^{Pl(H)} {\frac{{E(D,D)}}
{{Pl(H) - Bel(H)}}dt}  \hfill \\
  \quad \quad  \;\; = \frac{{E(D,H) - E(D,D)}}
{{Pl(H) - Bel(H)}} \cdot \left. \frac{{t^2 }}
{2} \right|_{Bel(H)}^{Pl(H)}  + \frac{{E(D,D)}}
{{Pl(H) - Bel(H)}} \cdot \left. t \right|_{Bel(H)}^{Pl(H)}  \hfill \\
  \quad \quad  \;\; = \frac{{E(D,H) - E(D,D)}}
{{Pl(H) - Bel(H)}}\left( {\frac{{\left[ {Pl(H)} \right]^2 }}
{2} - \frac{{\left[ {Bel(H)} \right]^2 }}
{2}} \right) + \frac{{E(D,D)}}
{{Pl(H) - Bel(H)}}\left[ {Pl(H) - Bel(H)} \right] \hfill \\
  \quad \quad  \;\; = \left[ {E(D,H) - E(D,D)} \right]\frac{{Pl(H) + Bel(H)}}
{2} + E(D,D) \hfill \\
  \quad \quad  \;\; = \left( {0 - \frac{V}
{2}} \right)\frac{{Pl(H) + Bel(H)}}
{2} + \frac{V}
{2} \hfill \\
  \quad \quad  \;\; = \frac{V}
{2} - \frac{V}
{2} \cdot \frac{{Pl(H) + Bel(H)}}
{2} \hfill \\.
\end{gathered}
\]

According to the Bishop-Canning theorem, the following condition is used to find the belief-based ESS $J$:
\begin{equation}
\mathbb{E}\left[ {H,J} \right] = \mathbb{E}\left[ {D,J} \right],
\end{equation}
Namely,
\[
V - \frac{{V + C}}
{2} \cdot \frac{{Pl(H) + Bel(H)}}
{2} = \frac{V}
{2} - \frac{V}
{2} \cdot \frac{{Pl(H) + Bel(H)}}
{2},
\]
\[
\frac{{Pl(H) + Bel(H)}}
{2} = \frac{V}
{C}.
\]
Hence,
\begin{equation}
\left\{ \begin{gathered}
  Bel(H) = \frac{V}
{C} - \delta  \hfill \\
  Pl(H) = \frac{V}
{C} + \delta  \hfill \\
\end{gathered}  \right.
\end{equation}
Formally, the belief-based ESS is shown as below.
\begin{equation}
J = tH + (1-t)D,  \qquad t \in [\frac{V}{C} - \delta, \quad \frac{V}{C} + \delta],
\end{equation}
where $0 \leqslant \frac{V}{C} - \delta  \leqslant \frac{V}{C} + \delta  \leqslant 1$ and $V < C$. Also, the belief-based ESS $J$ can represented as the forms of belief function,
\begin{equation}\label{BeliefStrategyJHDGame}
\left\{ \begin{array}{l}
 J(H) = \frac{V}{C} - \delta, \\
 J(D) = 1 - \frac{V}{C} - \delta, \\
 J(H,D) = 2\delta. \\
 \end{array} \right.
\end{equation}

\begin{figure}[htbp]
\begin{center}
\psfig{file=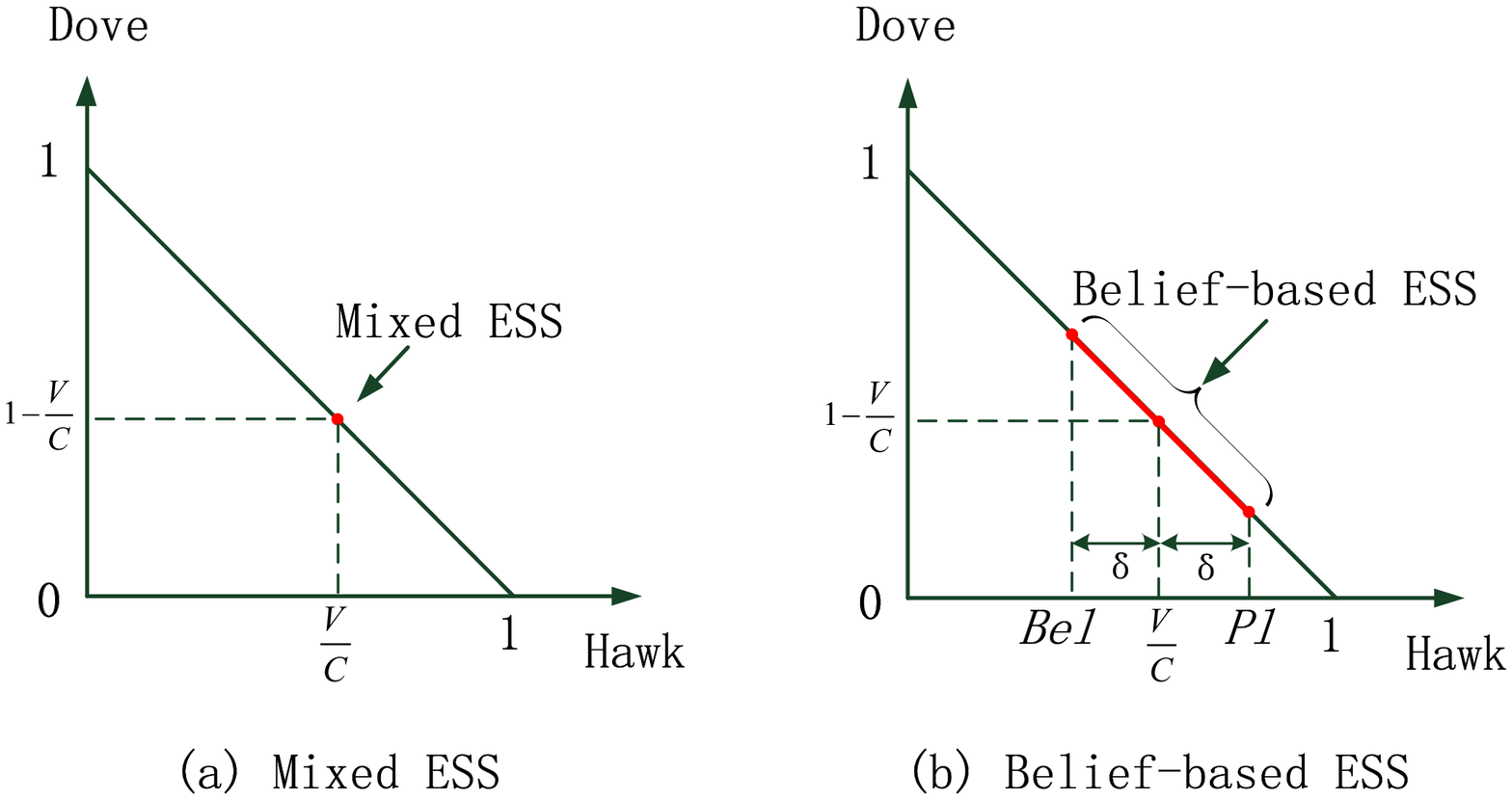,scale=0.55,angle=-90}
\caption{Graphical representation of mixed ESS and belief-based ESS}\label{MixedBelief}
\end{center}
\end{figure}

Figure \ref{MixedBelief} features the mixed ESS and belief-based ESS in the Hawk-Dove game. It is explicit that in a game with two pure strategies the mixed ESS is a point, while the belief-based ESS is a segment determined by parameter $\delta$ geometrically. When $\delta = 0$, the belief-based ESS is totally reduced to the mixed ESS. The parameter $\delta$ is a measure to reflect the uncertainty of belief strategy or belief-based ESS.

In terms of the Bishop-Canning theorem, $\mathbb{E}\left[ {H,J} \right] = \mathbb{E}\left[ {D,J} \right] = \mathbb{E}\left[ {J,J} \right]$. In order to further verify the stability of $J$ against invasion, the condition given in Eq.(\ref{BESScondition2}) is examined.
\[
\begin{gathered}
  \mathbb{E}\left[ {J,H} \right] = \int_{ - \infty }^{ + \infty } {\left[ {tE(H,H) + (1 - t)E(D,H)} \right]f(t)dt}  \hfill \\
  \quad \quad  \;\; = \int_{Bel(H)}^{Pl(H)} {\frac{{tE(H,H) + (1 - t)E(D,H)}}
{{Pl(H) - Bel(H)}}dt}  \hfill \\
  \quad \quad  \;\; = \int_{Bel(H)}^{Pl(H)} {\frac{{t\left[ {E(H,H) - E(D,H)} \right]}}
{{Pl(H) - Bel(H)}}dt}  + \int_{Bel(H)}^{Pl(H)} {\frac{{E(D,H)}}
{{Pl(H) - Bel(H)}}dt}  \hfill \\
  \quad \quad  \;\; = \frac{{E(H,H) - E(D,H)}}
{{Pl(H) - Bel(H)}} \cdot \left. \frac{{t^2 }}
{2} \right|_{Bel(H)}^{Pl(H)}  + \frac{{E(D,H)}}
{{Pl(H) - Bel(H)}} \cdot \left. t \right|_{Bel(H)}^{Pl(H)}  \hfill \\
  \quad \quad  \;\; = \frac{{E(H,H) - E(D,H)}}
{{Pl(H) - Bel(H)}}\left( {\frac{{\left[ {Pl(H)} \right]^2 }}
{2} - \frac{{\left[ {Bel(H)} \right]^2 }}
{2}} \right) + \frac{{E(D,H)}}
{{Pl(H) - Bel(H)}}\left[ {Pl(H) - Bel(H)} \right] \hfill \\
  \quad \quad  \;\; = \left[ {E(H,H) - E(D,H)} \right]\frac{{Pl(H) + Bel(H)}}
{2} + E(D,H) \hfill \\
  \quad \quad  \;\; = \left( {\frac{{V - C}}
{2} - 0} \right)\frac{{(V/C + \delta ) + (V/C - \delta )}}
{2} + 0 \hfill \\
  \quad \quad  \;\; = \frac{{V - C}}
{2} \cdot \frac{V}
{C} \hfill \\,
\end{gathered}
\]
\[
\begin{gathered}
  \mathbb{E}\left[ {J,D} \right] = \int_{ - \infty }^{ + \infty } {\left[ {tE(H,D) + (1 - t)E(D,D)} \right]f(t)dt}  \hfill \\
  \quad \quad  \;\; = \int_{Bel(H)}^{Pl(H)} {\frac{{tE(H,D) + (1 - t)E(D,D)}}
{{Pl(H) - Bel(H)}}dt}  \hfill \\
  \quad \quad  \;\; = \int_{Bel(H)}^{Pl(H)} {\frac{{t\left[ {E(H,D) - E(D,D)} \right]}}
{{Pl(H) - Bel(H)}}dt}  + \int_{Bel(H)}^{Pl(H)} {\frac{{E(D,D)}}
{{Pl(H) - Bel(H)}}dt}  \hfill \\
  \quad \quad  \;\; = \frac{{E(H,D) - E(D,D)}}
{{Pl(H) - Bel(H)}} \cdot \left. \frac{{t^2 }}
{2} \right|_{Bel(H)}^{Pl(H)}  + \frac{{E(D,D)}}
{{Pl(H) - Bel(H)}} \cdot \left. t \right|_{Bel(H)}^{Pl(H)}  \hfill \\
  \quad \quad  \;\; = \frac{{E(H,D) - E(D,D)}}
{{Pl(H) - Bel(H)}}\left( {\frac{{\left[ {Pl(H)} \right]^2 }}
{2} - \frac{{\left[ {Bel(H)} \right]^2 }}
{2}} \right) + \frac{{E(D,D)}}
{{Pl(H) - Bel(H)}}\left[ {Pl(H) - Bel(H)} \right] \hfill \\
  \quad \quad  \;\; = \left[ {E(H,D) - E(D,D)} \right]\frac{{Pl(H) + Bel(H)}}
{2} + E(D,D) \hfill \\
  \quad \quad  \;\; = \left(V - \frac{V}{2} \right) \cdot \frac{{(V/C + \delta ) + (V/C - \delta )}}
{2} + \frac{V}{2} \hfill \\
  \quad \quad  \;\; = \frac{{V}}
{2} \cdot \frac{V}
{C}  + \frac{V}{2} \hfill \\,
\end{gathered}
\]

\[
\mathbb{E}\left[ {H,H} \right] = \frac{{V - C}}{2}, \qquad and \qquad \mathbb{E}\left[ {D,D} \right] = \frac{{V}}{2}. \qquad \qquad \qquad \qquad \qquad \qquad
\]

It can be found $\mathbb{E}\left[ {J,H} \right] > \mathbb{E}\left[ {H,H} \right]$, and $\mathbb{E}\left[ {J,D} \right] > \mathbb{E}\left[ {D,D} \right]$,  when $V < C$. Hence, these formulas prove that the belief-based ESS $J$ is stable against invasion.

\section{Conclusions}\label{Sect4}
In short, we have reviewed the concept of ESS and the uncertainty involved in the mixed strategy. This uncertainty  mainly comes from the disturbance of the selection probability of pure strategies. In order to reflect such uncertainty, a belief strategy has been proposed based on Dempster-Shafer evidence theory. The proposed belief strategy is a generalization of mixed strategy. If the set of supports of a belief strategy only consists of single pure strategy, the belief strategy can reduce to a mixed strategy. What's more, on the basis of the belief strategy, a belief-based ESS is proposed, which, to large extent, extends the mixed ESS. The proposed belief strategy and belief-based ESS can provide more powerful tools to describe complicated interaction among agents.

\section*{Acknowledgements}
The work is partially supported by National Natural Science Foundation of China (Grant No. 61174022), Specialized Research Fund for the Doctoral Program of Higher Education (Grant No. 20131102130002), R\&D Program of China (2012BAH07B01), National High Technology Research and Development Program of China (863 Program) (Grant No. 2013AA013801), the open funding project of State Key Laboratory of Virtual Reality Technology and Systems, Beihang University (Grant No.BUAA-VR-14KF-02), Fundamental Research Funds for the Central Universities (Grant No. XDJK2014D034).





\bibliographystyle{elsarticle-num}
\bibliography{references}







\end{document}